\documentclass{jetplhome}
\twocolumn
\lat

\usepackage{epsfig}
\usepackage{color}

\title{ Fermi points and the Nambu sum rule in the polar phase of $^3$He }

\rtitle{Fermi points and the Nambu sum rule in the polar phase of $^3$He}

\sodtitle{Fermi points and the Nambu sum rule in the polar phase of $^3$He}

\author{M.\,A.\,Zubkov$^{+*}$\/\thanks{e-mail: zubkov@itep.ru}}

\rauthor{M.\,A.\,Zubkov }

\sodauthor{Zubkov}

\address{$^+$ Institute for Theoretical and Experimental Physics, B. Cheremushkinskaya 25, Moscow, 117259, Russia \\~\\
$^*$ Moscow Institute of Physics and Technology, 9, Institutskii per., Dolgoprudny, Moscow Region, 141700, Russia}

\date{\today}

\abstract{We discuss the polar phase of $^3$He, which is realized in the anisotropic aerogel. We consider it in the framework of the BCS model. In the absence of the spin - orbit interaction this model predicts the appearance of the Fermi line. However, it is topologically unstable. We demonstrate, that the spin - orbit interaction gives rise to the appearance of the two Fermi points instead of the Fermi line. In addition to the gapless Nambu - Goldstone bosons in this system the collective gapped bosonic states exist. Their gaps are calculated, and the corresponding Nambu sum rule is established. }


\begin{document}

\maketitle

\section{Introduction}

Polar phase of superfluid  $^3$He has been recently observed in strongly
anisotropic alumina aerogel \cite{Dmitriev2012,Dmitriev2014, Dmitriev2,Mineev2014}. The possibility that this phase  may appear in anisotropic aerogel has been proposed earlier in \cite{aerogel0}. The superfluid phases of $^3$He were widely discussed (see \cite{Volovik2003}, \cite{VollhardtWolfle1990} and references therein). The BCS theory of $^3$He based on the functional integral representation has been developed in a number of publications (see, for example, \cite{He3},\cite{He3gauss}, \cite{He3B}, \cite{BrusovPopov1980}) and was summarized recently, for example, in \cite{Brusovs}.

One of the most interesting properties of the superfluid $^3$He is that it is able to simulate the phenomena typical for the high energy physics theory. In particular, the emergent relativistic Weyl fermions appear naturally in the A phase \cite{Volovik2003}, \cite{VolovikVachaspati1996}, \cite{VolovikKhazan1982}, \cite{Volovik1990}. This possibility is related to the appearance of the Fermi points. At the same time, it is generally believed, that in the polar phase the Fermi line appears \cite{Volovik2003}.  In the present paper we demonstrate that it disappears if we take into account the spin - orbit interaction. Instead of it the two Fermi points appear, which allows, in principle, to simulate relativistic physics in the $^3$He inside aerogel.

Collective modes in superfluids possess an analogy to the Higgs modes of particle physics. In $^3$He - B \cite{Nambu1985}, and later in $^3$He - A, and in the 2D phases of $^3$He the Nambu sum rule was established that relates the energy gaps of the bosonic modes and the fermion "gap" \cite{VZ2015}, \cite{VolovikZubkov2014}, \cite{VolovikZubkovHiggs}. The Nambu sum rule also takes place in a number of relativistic models of the Nambu - Jona - Lasinio type \cite{VolovikZubkovHiggs}. In the present paper we analyse the spectrum of  bosonic collective modes in the polar phase and demonstrate, that here  a kind of the Nambu sum rule exists as well.

\section{The BCS model of $^3$He}


According to \cite{He3} Helium - 3 may be described by the effective theory with the partition function
\begin{eqnarray}
Z & = & \int D \Psi \, D \bar{A} D A\,  {\rm exp}\Big(\frac{1}{g}\sum_{p, i, \alpha} \bar{A}_{i, \alpha}(p) A_{i, \alpha}(p) \nonumber\\ &&- \sum_{p_1 p_2} \bar{\Psi}_{p_1}  G^{-1}_{p_1 p_2} \Psi_{p_2} \Big)
\end{eqnarray}
where $\Psi$ is the 4 - component Nambu - Gorkov spinor
$\Psi_p = \left( \begin{array}{cccc} a_+(p) & a_-(p) & \bar{a}_-(-p) & -\bar{a}_+(-p) \end{array}\right)^T$
, $A_{i, \alpha}$ is the auxiliary bosonic field,
while $G$ is the quasiparticle Green function. It may be represented as follows
\begin{eqnarray}
G^{-1}_{p_1 p_2} &=& (i\omega - v_F (|k|-k_F) \gamma^5 )\delta_{p_1 p_2}\nonumber\\&&+ \gamma^\alpha \, A_{\alpha i}(p_1-p_2)  \, \frac{\hat{k}_1^i+\hat k_2^i}{\sqrt{\beta V}} \frac{1-\gamma^5}{2}\nonumber\\&&- \gamma^\alpha \, \bar{A}_{\alpha i}(p_2-p_1)  \, \frac{\hat{k}_1^i+\hat k_2^i}{\sqrt{\beta V}} \frac{1+\gamma^5}{2} \label{matrix22}
\end{eqnarray}
where $\gamma^k$ are the Gamma - matrices in chiral representation, while $p = (\omega, k), \quad \hat{k} = \frac{k}{|k|}$. Here $V$ is the $3D$ volume, while $\beta = 1/T$ is the imaginary time extent of the model (i.e. the inverse temperature). Both $\beta$ and $V$ should be set to infinity at the end of the calculations.   $a_{\pm}(p)$ is the fermion Grassmann variable in momentum space, $v_F$ is Fermi velocity, $k_F$ is Fermi momentum, $g$ is the effective coupling constant. The spin-orbit coupling in liquid $^3$He (the dipole-dipole interaction) is relatively small, and the spin and orbital rotation groups, $SO_3^{ S}$ and $SO_3^{L}$, can be considered independently in first approximation.
If we neglect the spin - orbit interactions, then the order parameter matrix $A_{\alpha i}$ in the polar phase vacuum has the
form:
\begin{equation}
A^{(0)}_{\alpha i}=(\beta V)^{1/2} \frac{\Delta}{2} \, \delta_{p0} ~e^{i\Phi}~\hat d_{\alpha} \hat m_{i}\,,
 \label{Apolar}
\end{equation}
where $\hat {\bf d}$ and $\hat {\bf m}$ are unit vectors. Phase $\Phi$ may be set to zero. The anisotropy of aerogel fixes the direction of $\hat{\bf m}$. We take this into account via the following term in the action
$$S_{m} = -\frac{1}{g_m} \sum_{p,i,\alpha}{\bar A}_{\alpha i}(p) \hat{m}^i A_{\alpha j}(p) {\hat m}^j$$
with the new coupling constant $g_m$. It appears, that only sufficiently strong interaction of this form allows to make vacuum of the polar phase stable.

The spin - orbit interaction gives the following contribution to the effective low energy action \cite{VollhardtWolfle1990}:
\begin{eqnarray}
S_{SO}[A] & = & \frac{3}{5}g_D \sum_{p}\bar{A}_{i, \alpha}(p) A_{j, \beta}(p)\Big(\delta_{i\alpha}\delta_{j\beta}+\delta_{j\alpha}\delta_{i\beta}\nonumber\\&&-\frac{2}{3}\delta_{ij}\delta_{\alpha\beta} \Big),\label{SSO}
\end{eqnarray}
where $g_D$ is the new coupling constant. If we substitute to this expression the field $A_{i,\beta}$ of Eq. (\ref{Apolar}), it gives
$S_{SO}[A]  =  \frac{3}{10}g_D \Delta^2 \Big((\hat d \hat m)^2 - \frac{1}{3}\Big) \beta V $. Minimum of this expression corresponds to vector $\hat d$ orthogonal to the anisotropy axis of aerogel (directed along $\hat m$). We will see, however, that in the presence of the spin - orbit interaction the gap equation does not admit the solution of the form of Eq. (\ref{Apolar}). This solution with $(\hat d \hat m)=0$ appears  only as the limiting case of the precise solution at $g_D \to 0$. 

\section{"Gap" equation and the fermionic excitations in the absence of spin - orbit interactions}

There are $18$ modes of the fluctuations $\delta A_{i \alpha}=A_{i \alpha}- A^{(0)}_{i \alpha}$ around the condensate. In the absence of spin - orbit interactions tensor $\delta A_{i\alpha}$ realizes the representation of the $SO_S(2)\otimes SO_L(2)$ symmetry group. The gap equation receives the form
$$
 \Big(\frac{1}{g}  - \frac{1}{g_m}\Big)    \hat{m}^i {\hat d}^\alpha  \Delta  = -2 \int \frac{d^3k d\omega}{(2\pi)^4}{\rm Tr} \gamma^5 \gamma^\alpha \hat{k}^i G(i\omega,k)
$$
with
\begin{equation}
G(\epsilon,k) = \Big({\sum_{\mu = 1,2,3,5}{\cal P}_\mu(\epsilon,k) \gamma^\mu - {\cal M}(k)}\Big)^{-1}\gamma^5\label{G1}
\end{equation}
and
$ {\cal P}^5 = \epsilon, \, {\cal P}^{\alpha} = \Delta {\hat d}^\alpha ({\hat m}{\hat k}), \, {\cal M} = v_F(|k|-k_F)$.
We may rewrite this equation as follows
\begin{eqnarray}
 \Big(\frac{1}{g}  - \frac{1}{g_m}\Big) &  = &   \frac{k_F^2}{\pi^2 v_F} \int  \frac{{\rm cos}^2\,\theta dt \,d \, {\rm cos} \, \theta \, }{ \sqrt{t-4 \Delta_\theta^2}\sqrt{t}}  \theta(t-4 \Delta^2_\theta) \nonumber\\ &\approx &
    \frac{k_F^2}{\pi^2 v_F}\Big(\frac{2}{3}\,{\rm log}\,\frac{4v_F^2 {\cal K}^2}{\Delta^2} + \frac{4}{9} \Big)\label{gapeq03}
\end{eqnarray}
where $\Delta_\theta = \Delta (\hat m \hat k) = \Delta \, {\rm cos}\,\theta$.
This equation relates the value of the "gap" $\Delta$ with the values of the coupling constants
and with the momentum cutoff ${\cal K}$. The integration over $t$ is in the ranges $0 < t < \Lambda^2_\theta = 4(v^2_F {\cal K}^2 + \Delta^2 {\rm cos}^2\, \theta)$. It corresponds to the integration over momenta $k_F - {\cal K} < |k| < k_F + {\cal K}$.
It is normally supposed, that
$
\Delta \ll v_F {\cal K} \ll v_F k_F
$.
In the presence of the condensate of the form of Eq. (\ref{Apolar}) the fermionic quasiparticles are described by the partition function
$Z = \int D \bar{a} D a \, {\rm exp}\Big( - \sum_p \bar{\Psi}_p  G^{-1}(i \omega, k) \Psi_p \Big)$
, where $G$ is the quasiparticle Green function given by Eq. (\ref{G1}), and $p = (\omega, k)$.
This gives dispersion that does not depend on spin as well as on $\hat d$ and $\Phi$:
$$
E(k) = \pm \sqrt{v_F^2(|k|-k_F)^2 + \Delta^2 (\hat m \hat{k})^2 }
$$
It has its zero along the Fermi line that is the intersection of the sphere $|k| = k_F$ and the plane $\hat m k = 0$.

In order to consider the question about the stability of the Fermi line let us consider the situation, when variations of the field $A$ around its condensate have the form
$A_{\alpha i}=(\beta V)^{1/2} \frac{1}{2} \, \delta_{p0} C_{\alpha i}$.
Then the fermion Green function is given by
$G^{-1}(p) = i\omega - v_F (|k|-k_F) \gamma^5 + \gamma^5 \gamma^\alpha \, {\rm Re}\, C_{\alpha i} \, \hat{k}^i + i \gamma^\alpha\, {\rm Im}\, C_{\alpha i}\, \hat{k}^i $.
One can easily see, that at $\omega = 0$ matrix $G^{-1}$ anti - commutes with the matrix of time reversal transformation ${\bf K} = \gamma^5 \gamma^4 $.
The  Fermi - line would be topologically stable, if the following invariant is nonzero
${\cal N}_2=  {\bf tr} \oint_C \frac{dl}{4\pi } [{\bf K} {\cal G}({\bf p}) \partial_l  {\cal G}^{-1}({\bf p})]$.
One can check, however, that for the contour $C$ winding around the Fermi line in the polar phase the value of ${\cal N}_2$ is equal to $0$, i.e. the contributions of the two components of spin compensate  each other, which means that the Fermi line in the polar phase is not stable with respect to the perturbations of the condensate.

\section{Polar phase with the spin - orbit interaction taken into account}


 Let us consider the condensate of the form
\begin{equation}
A^{(0)}_{\alpha i}=(\beta V)^{1/2} \frac{\Delta}{2} \, \delta_{p0} \Big( \hat d^\alpha \hat m^i + \kappa^{\alpha i}\Big)\,,
 \label{Apolar3}
\end{equation}
with $|\kappa^{\alpha i}|\ll 1$. The gap equation receives the form
\begin{eqnarray}
&& \Big(\frac{1}{g}  - \frac{1}{g_m}\Big)    \kappa^{\alpha i}  \Delta \nonumber\\ && + \Big(\frac{1}{g}  - \frac{1}{g_m}-\frac{2}{5}g_D\Big)    \hat{m}^i {\hat d}^\alpha  \Delta \nonumber\\&& + \frac{3}{5}g_D \hat m^\alpha \hat d^i \Delta =- 2\int \frac{d^3k d\omega}{(2\pi)^4}{\rm Tr} \gamma^5 \gamma^\alpha \hat{k}^i G(i\omega,k)
\end{eqnarray}
with the Green function of the form of Eq. (\ref{G1}) with
$ {\cal P}^5 = \epsilon, \, {\cal P}^{\alpha} = \Delta \Big({\hat d}^\alpha {\hat m}^i + \kappa^{\alpha i} \Big){\hat k}^i, \, {\cal M} = v_F(|k|-k_F)$.
(It is taken into account that $\hat m \perp \hat d$.)
In this equation we keep the terms linear in $g_D$ and $\kappa^{\alpha i}$. This allows to derive expression for the correction to the condensate:
\begin{equation}
\kappa^{\alpha i}    = a\, {\hat d}^\alpha {\hat m}^i  + b\,  {\hat d}^i {\hat m}^\alpha\label{kappa}
\end{equation}
with $
a = - \frac{1}{5} \frac{g_D}{ \tilde J^{(1)}} \approx \frac{3v_F \pi^2}{10} \frac{g_D}{k_F^2}, \,\quad b = \frac{6}{5}\frac{g_D}{J^{(0)}-3J^{(1)}} \approx \frac{v_F 9\pi^2}{20} \frac{g_D}{k_F^2}
$. Here
$J^{(0)}  =  \frac{1}{4 \pi^2 v^3_F} \int d \, {\rm cos} \, \theta \,\int_{4\Delta^2_\theta}^{\Lambda^2_\theta} dt \frac{t - 4 \Delta_\theta^2 + 4 v_F^2 k_F^2}{ \sqrt{t-4 \Delta_\theta^2}\sqrt{t}} $, $J^{(1)}  =  \frac{1}{4 \pi^2 v^3_F} \int d \, {\rm cos} \, \theta \,\int_{4\Delta^2_\theta}^{\Lambda^2_\theta} dt \frac{t - 4 \Delta_\theta^2 + 4 v_F^2 k_F^2}{ \sqrt{t-4 \Delta_\theta^2}\sqrt{t}} (\hat k \hat m)^2 $, $\tilde J^{(1)}  =  \Delta^2 \frac{\partial}{\partial \Delta^2} J^{(1)}$.
In the presence of the condensate of the form of Eq. (\ref{Apolar3}) we have the fermion Green function of the form of Eq. (\ref{G1}) with
$ {\cal P}^5 = \epsilon, \, {\cal P}^{\alpha} = \Delta \Big((1+a)\, {\hat d}^\alpha {\hat m}^i  + b\,  {\hat d}^i {\hat m}^\alpha \Big){\hat k}^i, \, {\cal M} = v_F(|k|-k_F)$.
This gives the dispersion:
\begin{equation}
E(k) = \pm \sqrt{v_F^2(|k|-k_F)^2 + \Delta^2 \Big((1+a)^2(\hat m \hat{k})^2 + b^2 (\hat d \hat{k})^2\Big) }\nonumber
\end{equation}
It has the two zeros at the intersection of the sphere $|k| = k_F$ and the planes $(\hat m k) = 0$ and $(\hat d k) = 0$. The two Fermi points appear:
${\bf K}^\pm = \pm k_F\, {\bf l}, \quad {\bf l} = [\hat m \times \hat d]$.
We denote $k = {\bf K}^\pm + q$ and make the transformation of spinors near ${\bf K}^+$: $\Psi_+ \to \gamma^0 \Psi_+$. As a result the action receives the form
$S_{{\bf K}^\pm}  =  \sum_q \bar{\Psi}_\pm \Gamma^0\Big(i\omega \Gamma^0  + (v_F (\hat l q)\Gamma^3 +  (1+a) \frac{\Delta}{k_F} \,  (\hat m q) \Gamma^1 +  b\, \frac{\Delta}{k_F} ( \hat d q) \Gamma^2 )\Big) \Psi_\pm$,
where we denoted
$\Gamma^0 = i\gamma^5 \gamma^4, \Gamma^3 = \Gamma^0\gamma^5, \, \Gamma^1 = \Gamma^0 \gamma^5(\gamma \hat d), \, \Gamma^2 = \Gamma^0 \gamma^5(\gamma \hat m)$.
Those matrices satisfy
$
\{\Gamma^i,\Gamma^j\}=-2\delta^{ij}
$ for $i,j=1,2,3$.
One can see, that the two Fermi points correspond to massless Dirac spinors. Fermi velocities of the quasiparticle excitations along different directions are different:
$
v_{1} = v_F \gg v_2 = (1+a) \frac{\Delta}{k_F} \gg v_3 = b \frac{\Delta}{k_F}$.
Those spinors have the Hamiltonian $ (v_1 \Gamma^1 (\hat l q) +  v_2 \Gamma^2 (\hat m q) + v_3 \Gamma^3 (\hat d q) )$ that commutes with matrix $\Gamma^5 = ({\hat l}\gamma) \gamma^4$.

\section{Bosonic collective excitations}

In our calculation for simplicity we neglect the spin - orbit interaction.
According to \cite{He3gauss,He3B} the quadratic part of the effective action for the fluctuations around the condensate has the form:
\begin{equation}
S^{(1)}_{eff} = (\bar{u},\bar{v}) [1/g - \Omega -  \Pi(E)]   \left(\begin{array}{c}{u}\\v \end{array}\right),
\end{equation}
where
$\Omega^{\alpha i}_{\bar{\alpha} \bar{i}} = \frac{1}{g_m} \delta^\alpha_{\bar \alpha}  \hat{m}^i {\hat m}^{\bar{i}}$
while $
u_{ i \alpha}(p) =  \frac{1}{2}\Big({\rm Re}\delta A_{i \alpha}(p) + {\rm Re}\delta A_{i \alpha}(-p)+i{\rm Im}\delta A_{i \alpha}(p) - i{\rm Im}\delta A_{i \alpha}(-p)\Big)
$ and
$v_{ i \alpha}(p)  =  \frac{1}{2}\Big({\rm Im}\delta A_{i \alpha}(p) + {\rm Im}\delta A_{i \alpha}(-p)-i{\rm Re}\delta A_{i \alpha}(p) + i{\rm Re}\delta A_{i \alpha}(-p)\Big)
$.
Here $E$ is the energy gap of the collective excitation. Its momentum is supposed to be equal to zero. The components of the polarization operator are given by
\begin{eqnarray}
\Big[\Pi^{\bar{u}u}(E)\Big]^{\alpha i}_{\bar \alpha \bar i} &=& i \int \frac{d^3 k d \epsilon}{(2\pi)^4} {\rm Tr}\,G(\epsilon,k)\gamma^5\gamma^{\alpha}\hat{k}^i\nonumber\\&&G(\epsilon-E,k)\gamma^5\gamma^{\bar\alpha}\hat{k}^{\bar i}
\end{eqnarray}
and
\begin{eqnarray}
\Big[\Pi^{\bar{v}v}(E)\Big]^{\alpha i}_{\bar \alpha \bar i} & = & -i \int \frac{d^3 k d \epsilon}{(2\pi)^4} {\rm Tr}\,G(\epsilon,k)\gamma^{\alpha}\hat{k}^i\nonumber\\&& G(\epsilon-E,k)\gamma^{\bar\alpha}\hat{k}^{\bar i}
\end{eqnarray}
that may be represented as
$\Pi(E) = \frac{1}{\pi}\int_{0}^{\infty} d z \frac{\rho(z)}{z-E^2}$.
The spectral function $\rho \sim \sum |F_{Q \rightarrow f f}|^2$, and $|F_{Q \rightarrow f f}|^2$ is the probability that the given mode $Q$ decays  into two fermions.
Here
\begin{eqnarray}
&&\Big[\rho^{\bar uu}\Big]^{\alpha i}_{\bar \alpha \bar i} = \frac{1}{4 \pi v^3_F} \int \frac{d\phi}{2 \pi} d \, {\rm cos} \, \theta \, \frac{t - 4 \Delta_\theta^2 + 4 v_F^2 k_F^2}{ \sqrt{t-4 \Delta_\theta^2}\sqrt{t}} \nonumber\\&&\Big( t \delta^{\alpha  \bar\alpha} -4 \Delta_\theta^2  {\hat d}^\alpha {\hat d}^{\bar{\alpha}}  \Big)\hat{k}_+^i \hat{k}_+^{\bar i} \theta(t-4 \Delta^2_\theta)\theta( \Lambda^2_\theta-t)
\end{eqnarray}
and
\begin{eqnarray}
&&\Big[\rho^{\bar vv}\Big]^{\alpha i}_{\bar \alpha \bar i} =  \frac{1}{4\pi v_F^3}\int \frac{ d\,\phi\,  d \, {\rm cos} \, \theta}{ 2\pi } \,\frac{t - 4 \Delta_\theta^2 + 4 v_F^2 k_F^2}{ \sqrt{t-4 \Delta_\theta^2}\sqrt{t}} \nonumber\\&&\Big( (t-4 \Delta^2_\theta) \delta^{\alpha  \bar\alpha} + 4 \Delta_\theta^2  {\hat d}^\alpha {\hat d}^{\bar{\alpha}}  \Big)\hat{k}_+^i \hat{k}_+^{\bar i}\theta(t-4 \Delta^2_\theta)\theta( \Lambda^2_\theta-t)
\end{eqnarray}
We denote
$
\hat{k}_+ = ({\rm sin}\,\theta\,{\rm cos}\, \phi, {\rm sin}\,\theta\,{\rm sin}\, \phi, {\rm cos}\,\theta)
$,
$ E =\sqrt{t}$,  and $\Delta_\theta \equiv \Delta (\hat{m}\hat{k}_+)\equiv \Delta \, {\rm cos} \,\theta$.

For the definiteness let us suppose that $\hat m$ is directed along the third axis while $\hat d$ is directed along the second axis. There should be $3$ Goldstone modes:
at $L=S=0$ - one $v$ mode, and at $L=0,S=1$ - two $u$ modes.
At the same time for $g_m \to \infty$ in the $L=1,S=0$ channel the $u$ mode is to be gapless as well. The energy gaps appear as the zeros of ${\rm Det}\,[1/g - \Omega -  \Pi(E)]$.  Notice, that since the fermions are gapless in the polar phase, the energy gaps of the collective bosonic modes contain imaginary parts, that correspond to their decays to fermions.

{$L=S=0$: we take components with $\alpha = 2$, $i=3$.}
In the $v$ - channel at $S=L=0$  the energy gap is equal to zero. The corresponding equation for the determination of $E$ at $E=0$ is equivalent to the gap equation.
In the $u$ channel we have the following equation for the determination of the energy gap:
\begin{eqnarray}
1/g - 1/g_m &=&  \int_{-1}^1 {\rm cos}^2 \, \theta d \, {\rm cos} \, \theta \int_{4\Delta^2_\theta}^{\Lambda_\theta^2}  {dt}
 \frac{  1  }{4 \pi^2 v^3_F }\nonumber\\&&\frac{t - 4 \Delta_\theta^2 + 4 v_F^2 k_F^2}{ \sqrt{t-4 \Delta_\theta^2}\sqrt{t}} \,\frac{ t-4 \Delta_\theta^2 }{t-E^2_{u,L=0,S=0}}\label{eq1}
\end{eqnarray}
Recall that $\Delta_\theta = \Delta {\rm cos} \theta$ while the energy cutoff $\Lambda_\theta$ and the momentum cutoff ${\cal K}$ are related by expression
$\Lambda^2_\theta /4 = v_F^2 {\cal K}^2 + \Delta_\theta^2$. Comparing Eq. (\ref{eq1}) with the gap equation we come to the following approximate expression for the real part of the energy gap
\begin{equation}
E_{u,S=0,L=0} \approx  \sqrt{ 4 \langle \Delta_\theta^2\rangle}  \end{equation}
where averaging is over the angles $\theta$. Here
$\langle \Delta_\theta^2\rangle = \frac{\int_{-\Delta}^\Delta d \Delta_\theta \,\Delta_\theta^4}{\int_{-\Delta}^\Delta d \Delta_\theta \,\Delta_\theta^2}$.
The estimate follows
$ E_{v,S=0,L=0} = 0, \, E_{u,S=0,L=0} \approx  \sqrt{ 12/5}  \Delta  $. 
This estimate may be checked via the numerical solution of Eq. (\ref{eq1}). The integrals in this equation may be taken and the result is expressed through the hypergeometric functions:
\begin{eqnarray}
0 & = & \frac{ 4 k^2_F  }{ \pi^2 v_F } \Big[\frac{1}{4}w^4\sqrt{\pi} \Big(\frac{3}{8w}\pi^{3/2}-\frac{32}{15\sqrt{\pi}} F^{1/2, 1, 3}_{3/2, 7/2} (-w^2)\Big)\nonumber\\&&-\frac{1}{4} w^4 \sqrt{\pi} \Big(\frac{1}{2w} \pi^{3/2}-\frac{8}{3\sqrt{\pi}} F^{1/2, 1, 2}_{3/2, 5/2} ( -w^2)\Big)+\frac{1}{3}w^2+\frac{1}{3} \Big]\nonumber
\end{eqnarray}
 where
 $
 w = \frac{-i E_{u,L=0,S=0}}{2\Delta}
 $.
Numerical solution of this equation gives
\begin{eqnarray}
E_{u,S=0,L=0} &=& \sqrt{12/5} \, \Big( 1.007853779-0.3828669418\,i\Big) \, \Delta\nonumber \end{eqnarray}
One can see, that our estimate of the real part of the energy gap is in fact rather precise.

{$L=0,S=1$:  we take components with $\alpha = 1,3$, $i=3$.}
Here the situation is inverse compared to that of the case with $L=0,S=0$.  Our estimate is, therefore,
$E_{u,S=1,L=0} = E_{v,S=0,L=0} = 0, E_{v,S=1,L=0}  = E_{u,S=0,L=0}$.

{$L=1,S=0$:  we take components with $\alpha = 2$, $i=1,2$.}
In the $u$ channel we have
\begin{eqnarray}
\frac{1}{g} &=&  \int_{-1}^1 \frac{1-{\rm cos}^2 \, \theta}{2} d \, {\rm cos} \, \theta \int_{4\Delta^2_\theta}^{\Lambda_\theta^2}  {dt}
\frac{  1  }{4 \pi^2 v^3_F }\nonumber\\ &&\frac{t - 4 \Delta_\theta^2 + 4 v_F^2 k_F^2}{ \sqrt{t-4 \Delta_\theta^2}\sqrt{t}} \,\frac{t-4 \Delta_\theta^2}{t-E^2_{u,L=1,S=0}}\label{eq1_}
\end{eqnarray}
It appears that at $E_{u,L=1,S=0}=0$ the right hand side of this equation is identical to that of the gap equation. In order to demonstrate this we may rewrite this equation in the form with the integration over $k$ instead of integration over $t$, and perform integration over angles. Therefore, in the absence of the extra interaction that stabilizes direction of $\hat m$ in this channel the Goldstone boson appears as it should.
At the same time in the presence of this extra interaction we have the following equation for the determination of $E_{v,L=1,S=0}$:
\begin{eqnarray}
\frac{1}{g} &=&  \int_{-1}^1 \frac{1-{\rm cos}^2 \, \theta}{2} d \, {\rm cos} \, \theta \int_{4\Delta^2_\theta}^{\Lambda_\theta^2}  dt
\frac{  1 }{4 \pi^2 v^3_F }\nonumber\\&& \frac{t - 4 \Delta_\theta^2 + 4 v_F^2 k_F^2}{ \sqrt{t-4 \Delta_\theta^2}\sqrt{t}} \,\frac{t}{t-E^2_{v,L=1,S=0}}\label{eq2_}
\end{eqnarray}
There appears the critical value $g_m^{(c)}$ equal to
\begin{eqnarray}
\frac{1}{g^{(c)}_m} &=&  \int_{-1}^1 \frac{1-{\rm cos}^2 \, \theta}{2} d \, {\rm cos} \, \theta \int_{4\Delta^2_\theta}^{\Lambda_\theta^2}  {dt}\nonumber\\&&
\frac{  1  }{4 \pi^2 v^3_F }\,\frac{t - 4 \Delta_\theta^2 + 4 v_F^2 k_F^2}{ \sqrt{t-4 \Delta_\theta^2}\sqrt{t}} \,\frac{ 4 \Delta_\theta^2 }{t }\nonumber\\
&=& \frac{ 4 k^2_F  }{3\pi^2 v_F }
\end{eqnarray}
For $1/g_m^{(c)} > 0 > 1/g_m $ Eq. (\ref{eq1_}) has the solution with imaginary $E_{u,L=1,S=0}$. For $0 = 1/g_m$ the solution with $E_{u,L=1,S=0}=0$ appears, while for  $0 < 1/g_m$ there are no solutions of this equation. Equation Eq. (\ref{eq2_}) for $1/g_m^{} < 1/g_m^{(c)}$ has the solution with imaginary $E_{v,L=1,S=0}$. For $1/g_m^{(c)} = 1/g_m$ the solution with $E_{v,L=1,S=0}=0$ appears, while for  $1/g_m^{(c)} < 1/g_m$ there are no solutions of this equation.

{$L=1,S=1$: we take components with $\alpha = 1,3$; $i=1,2$.}
Our analysis shows that 
$E_{u,S=1,L=1} = E_{v,S=0,L=1},  E_{v,S=1,L=1} = E_{u,S=0,L=1}$.
In particular, in the limit $1/g_m \to 0$ the $4$ extra gapless fermions appear in the one loop approximation.


\section{Conclusions}

To conclude, we considered the polar phase of $^3$He (realized in the anisotropic aerogel) in the BCS model. The latter is discussed in the path integral formulation. The extra interaction that stabilizes the direction of vector $\hat m$ along the anisotropy axis of aerogel is added to this model. Without this interaction the polar phase is unstable, and it becomes stable only when this extra interaction is sufficiently strong.
Without the spin - orbit interaction there should be $3$ Goldstone bosons. Direct calculation of energy gaps confirms this expectation. Our analysis of the bosonic excitations demonstrates that in the channels with vanishing orbital momentum there are gapped modes. We calculated both real and imaginary parts of those gaps. The imaginary parts appear because the fermions are gapless, and those bosonic states may decay into them. The real parts are given by
$E_{v,S=0,L=0} = E_{u,S=1,L=0} = 0$, $E_{u,S=0,L=0} = E_{v,S=1,L=0} \approx \sqrt{12/5} \Delta$.
Correspondingly, there is the Nambu sum rule satisfied by these gaps:
\begin{eqnarray}
&& E^2_u+ E_v^2 \approx 4 \langle \Delta_\theta^2\rangle = \frac{12}{5} \Delta^2
\end{eqnarray}
It appears that when the interaction stabilizing the direction of $\hat m$ is sufficiently strong, and vacuum is stable, our one loop results do not predict the appearance of the Higgs modes with $L=1$. The mentioned values of the energy gaps were estimated neglecting the spin orbit interaction that keeps the direction of $\hat d$ in the plane orthogonal to $\hat m$. It is worth mentioning, that the corresponding part of the action should give rise to the energy gap of one of the would be Goldtone bosons with $L=0,S=1$. It should be equal to the so - called Leggett frequency of the polar phase \cite{VZ2015}. We do not consider the effect of spin orbit interaction on the spectrum of bosonic collective modes in the present paper.

However, we discuss in details the effect of the spin - orbit interaction on the fermionic quasiparticles. Without the spin - orbit interaction the considered BCS model predicts the appearance of the Fermi lines, along which the quasiparticle energy vanishes. However, the smooth deformation of the system is able to eliminate those Fermi lines. Indeed, this occurs due to the spin - orbit interaction. We show, that it changes the "gap" equations in such a way, that the conventional expression for the vacuum value of the order parameter Eq. (\ref{Apolar}) is already not the solution of the "gap" equation. Instead, the corrected form of the condensate (Eqs. (\ref{Apolar3}), (\ref{kappa})) appears. This form of the condensate gives rise to the two Fermi points  $\pm k_F \, \hat m \times \hat d$ instead of the Fermi line. Close to those two Fermi points the emergent Dirac fermions apppear. In turn, those Fermi points are not protected by topology, and the next corrections are able to eliminate them, so that small effective mass of the Dirac fermions might appear.


For the details of the calculation the author advises to consult the supplemental material to the present article.
The author kindly acknowledges useful discussions with G.E.Volovik. The present work was supported by Russian Science Foundation Grant No 16-12-10059.

\onecolumn

\begin{centerline}
{\Large Supplemental Material for}
\end{centerline}
\begin{centerline}
{\Large "Fermi points and the Nambu sum rule in the polar phase of $^3$He " }
\end{centerline}

\section*{"Gap" equation for the polar phase with the spin - orbit interaction taken into account}

In the presence of spin - orbit interactions we consider the condensate of the form
\begin{equation}
A^{(0)}_{\alpha i}=(\beta V)^{1/2} \frac{\Delta}{2} \, \delta_{p0} \Big( \hat d^\alpha \hat m^i + \kappa^{\alpha i}\Big)\,,
 \label{Apolar3}
\end{equation}
with $|\kappa^{\alpha i}|\ll 1$. The gap equation receives the form
\begin{eqnarray}
&&\Omega^{i\alpha}\equiv \Big(\frac{1}{g}  - \frac{1}{g_m}\Big)    \kappa^{\alpha i}  \Delta + \Big(\frac{1}{g}  - \frac{1}{g_m}-\frac{2}{5}g_D\Big)    \hat{m}^i {\hat d}^\alpha  \Delta \nonumber\\ && + \frac{3}{5}g_D \hat m^\alpha \hat d^i \Delta =- 2\int \frac{d^3k d\omega}{(2\pi)^4}{\rm Tr} \gamma^5 \gamma^\alpha \hat{k}^i G(i\omega,k)
\end{eqnarray}
with
\begin{equation}
G(\epsilon,k) = \frac{1}{\sum_{\mu = 1,2,3,5}{\cal P}_\mu(\epsilon,k) \gamma^\mu - {\cal M}(k)}\gamma^5
\end{equation}
and
$$ {\cal P}^5 = \epsilon, \, {\cal P}^{\alpha} = \Delta \Big({\hat d}^\alpha {\hat m}^i + \kappa^{\alpha i} \Big){\hat k}^i, \, {\cal M} = v_F(|k|-k_F)$$
(It is taken into account that $(\hat m \hat d)=0$.)
We may rewrite this equation as follows
\begin{equation}
\Omega^{i\alpha}=- 2\int \frac{d^3k d\omega}{(2\pi)^4}\frac{{\rm Tr} \Big({\cal P}_\mu(\epsilon,k) \gamma^\mu + {\cal M}(k)\Big) \gamma^\alpha \hat{k}^i }{\omega^2 + \Delta_\theta^2 + {\cal M}^2(k)}\label{gapeq00}
\end{equation}
where $\Delta_\theta = \Delta (\hat m \hat k)$.
Now
we have
\begin{eqnarray}
&& \Big(\frac{1}{g}  - \frac{1}{g_m}\Big)    \kappa^{\alpha i}  \Delta
= \frac{2}{5}g_D   \hat{m}^i {\hat d}^\alpha  \Delta - \frac{3}{5}g_D \hat m^\alpha \hat d^i \Delta  \nonumber\\
 &&+ {\kappa^{\alpha i}}\Delta \Big(\frac{1}{2}J^{(0)}-\frac{1}{2}J^{(1)}\Big)
 + ({\kappa^{\alpha j}} \hat m^j) \hat m^i\Delta \Big(\frac{3}{2}J^{(1)}-\frac{1}{2}J^{(0)}\Big)\nonumber\\
&&+(2 \kappa^{\beta i}\hat d^\beta)\hat d^\alpha \Delta \Big(\frac{1}{2}\tilde{J}^{(0)}-\frac{1}{2}\tilde{J}^{(1)}\Big)\nonumber\\&&+(2 \kappa^{\beta j}\hat d^\beta  \hat m^j)\hat d^\alpha \hat m^i\Delta \Big(\frac{3}{2}\tilde{J}^{(1)}-\frac{1}{2}\tilde{J}^{(0)}\Big)
\end{eqnarray}
where
\begin{eqnarray}
J^{(0)} & = & \frac{1}{4 \pi^2 v^3_F} \int \frac{d\phi}{2 \pi} d \, {\rm cos} \, \theta \int_{4\Delta_\theta^2}^{\Lambda^2_\theta} dt \, \frac{t - 4 \Delta_\theta^2 + 4 v_F^2 k_F^2}{ \sqrt{t-4 \Delta_\theta^2}\sqrt{t}} \nonumber\\
J^{(1)} & = & \frac{1}{4 \pi^2 v^3_F} \int \frac{d\phi}{2 \pi} d \, {\rm cos} \, \theta \, \int_{4\Delta_\theta^2}^{\Lambda^2_\theta} dt \frac{t - 4 \Delta_\theta^2 + 4 v_F^2 k_F^2}{ \sqrt{t-4 \Delta_\theta^2}\sqrt{t}} (\hat k \hat m)^2 \nonumber\\
\tilde J^{(0)} & = & \Delta^2 \frac{\partial}{\partial \Delta^2}J^{(0)} \nonumber\\
\tilde J^{(1)} & = & \Delta^2 \frac{\partial}{\partial \Delta^2} J^{(1)}
\end{eqnarray}
Here the energy cutoff $\Lambda_\theta$ and the momentum cutoff ${\cal K}$ are related by expression
$\Lambda^2_\theta /4 = v_F^2 {\cal K}^2 + \Delta_\theta^2$ (integration is over momenta with $|k-k_F|< {\cal K}$).
We keep the terms linear in $g_D$ and $\kappa^{\alpha i}$.
Here
\begin{eqnarray}
J^{(0)} &\approx &\frac{4 k_F^2}{\pi^2 v_F}\,\Big({\rm log}\,\frac{2 v_F {\cal K}}{\Delta} +1\Big)\nonumber\\
J^{(1)} &=&\frac{1}{g}-\frac{1}{g_m}\approx \frac{4 k_F^2}{3\pi^2 v_F}\,\Big({\rm log}\,\frac{2 v_F {\cal K}}{\Delta} +\frac{1}{3}\Big)\nonumber\\
\tilde J^{(1)} & = & -\frac{2 k_F^2}{3 \pi^2 v_F}
\end{eqnarray}
and
\begin{equation}
\kappa^{\alpha i}    = a \hat d^\alpha \hat m^i  + b \hat m^\alpha \hat d^i \label{kappa}
\end{equation}
with
$
a= \frac{3v_F \pi^2}{10} \frac{g_D}{k_F^2}
$
and
$
b = \frac{v_F 9\pi^2}{20} \frac{g_D}{k_F^2}
$.

\section*{Bosonic collective modes in the polar phase}

Let us calculate the energy gaps of the bosonic collective modes. In our calculation for simplicity we neglect spin - orbit interaction.
The quadratic part of the effective action for the fluctuations around the condensate has the form:
\begin{equation}
S^{(1)}_{eff} = (\bar{u},\bar{v}) [1/g - \Omega -  \Pi]   \left(\begin{array}{c}{u}\\v \end{array}\right),
\end{equation}
where
$$\Omega^{\alpha i}_{\bar{\alpha} \bar{i}} = \frac{1}{g_m} \delta^\alpha_{\bar \alpha}  \hat{m}^i {\hat m}^{\bar{i}}$$
while
$$
u_{ i \alpha}(p) = \frac{\delta A_{i \alpha}(p) +   \delta \bar A_{i \alpha}(-p)}{2}
$$
and
$$
v_{ i \alpha}(p) = \frac{\delta A_{i \alpha}(p) - \delta \bar A_{i \alpha}(-p)}{2i}
$$
Here
\begin{equation}
\Big[\Pi^{\bar uu}(E)\Big]^{\alpha i}_{\bar \alpha \bar i} = i \int \frac{d^3 k d \epsilon}{(2\pi)^4} {\rm Tr}\,G(\epsilon,k)\gamma^5\gamma^{\alpha}\hat{k}^iG(\epsilon-E,k)\gamma^5\gamma^{\bar\alpha}\hat{k}^{\bar i}
\end{equation}
and
\begin{equation}
\Big[\Pi^{\bar vv}(E)\Big]^{\alpha i}_{\bar \alpha \bar i} = -i \int \frac{d^3 k d \epsilon}{(2\pi)^4} {\rm Tr}\,G(\epsilon,k)\gamma^{\alpha}\hat{k}^iG(\epsilon-E,k)\gamma^{\bar\alpha}\hat{k}^{\bar i}
\end{equation}

The polarization operator can be represented as
\begin{equation}
\Pi(E) = \frac{1}{\pi}\int_{0}^{\infty} d z \frac{\rho(z)}{z-E^2}, \label{disp}
\end{equation}
where the spectral function may be calculated using the Cutkosky rule (see the Landau - Lifshitz course of theoretical physics, vol. 4, chapter 115)
\begin{eqnarray}
&& 2\Big[\rho^{\bar uu}\Big]^{\alpha i}_{\bar \alpha \bar i} = - 4 \pi^2  \int_{\epsilon>0} \frac{d^3 k d \epsilon}{(2\pi)^4} {\rm Tr}\,\Big({\cal P}_\mu(\epsilon,k) \gamma^\mu + {\cal M}(k)\Big)\nonumber\\&& \gamma^{\alpha}\hat{k}^i \Big({\cal P}_\mu(\epsilon-E,k) \gamma^\mu + {\cal M}(k)\Big)\gamma^{\bar\alpha}\hat{k}^{\bar i}\nonumber\\&&\delta({\cal P}^2(\epsilon,k) - {\cal M}^2(k))\delta({\cal P}^2(\epsilon-E,k) - {\cal M}^2(k))\nonumber\\
&=& -\sum_\pm \int \frac{ d\phi  \Big(k_F \pm \frac{\sqrt{t-4 \Delta^2_\theta}}{ 2 v_F}\Big)^2 d \, {\rm cos} \, \theta}{2 \pi 2 \pi v_F \sqrt{t - 4\Delta^2_\theta} \sqrt{t}} \,\nonumber\\&&\Big( (\frac{t}{2} - \Delta^2_\theta){\rm Tr} \gamma^{\alpha}\hat{k}_\pm^i  \gamma^{\bar\alpha}\hat{k}_\pm^{\bar i} + \Delta_\theta^2 {\rm Tr}(\hat d \gamma) \gamma^{\alpha}\hat{k}_\pm^i  (\hat d \gamma) \gamma^{\bar\alpha}\hat{k}_\pm^{\bar i} \Big)\nonumber\\&&\theta(t-4 \Delta^2_\theta)\theta(\Lambda^2_\theta-t)\nonumber
\\ &=& \frac{1}{2 \pi v^3_F} \int \frac{d\phi}{2 \pi} d \, {\rm cos} \, \theta \, \frac{t - 4 \Delta_\theta^2 + 4 v_F^2 k_F^2}{ \sqrt{t-4 \Delta_\theta^2}\sqrt{t}} \nonumber\\&&\Big( t \delta^{\alpha  \bar\alpha} -4 \Delta_\theta^2  {\hat d}^\alpha {\hat d}^{\bar{\alpha}}  \Big)\hat{k}_+^i \hat{k}_+^{\bar i} \theta(t-4 \Delta^2_\theta)\theta(\Lambda^2_\theta-t)
\end{eqnarray}
where
$
\hat{k} = ({\rm sin}\,\theta\,{\rm cos}\, \phi, {\rm sin}\,\theta\,{\rm sin}\, \phi, {\rm cos}\,\theta)
$
while
$ E/2 =\sqrt{t}/2 = \epsilon_+ =  \epsilon_- ; \, k_\pm =  k_F \pm \frac{\sqrt{t-4 \Delta_\theta}}{2 v_F}$, and $\Delta_\theta \equiv \Delta (\hat{m}\hat{k}_+)\equiv \Delta \, {\rm cos} \,\theta$. In the similar way
\begin{eqnarray}
&& 2\Big[\rho^{\bar vv}\Big]^{\alpha i}_{\bar \alpha \bar i}
= \frac{1}{2\pi v_F^3}\int \frac{ d\,\phi\,  d \, {\rm cos} \, \theta}{ 2\pi } \,\frac{t - 4 \Delta_\theta^2 + 4 v_F^2 k_F^2}{ \sqrt{t-4 \Delta_\theta^2}\sqrt{t}} \nonumber\\ &&\Big( (t-4 \Delta^2_\theta) \delta^{\alpha  \bar\alpha} + 4 \Delta_\theta^2  {\hat d}^\alpha {\hat d}^{\bar{\alpha}}  \Big)\hat{k}_+^i \hat{k}_+^{\bar i}\theta(t-4 \Delta^2_\theta)\theta(\Lambda^2_\theta-t)
\end{eqnarray}

\section*{Energy gaps and the Nambu sum rule}

Let us come to the evaluation of the energy gaps.

{$L=S=0$. We take components with $\alpha = 2$, $i=3$.}
In the $v$ - channel at $S=L=0$  the energy gap is equal to zero that leads to  the condition
\begin{eqnarray}
1/g - 1/g_m &=&
\int_{-1}^1 {\rm cos}^2 \, \theta d \, {\rm cos} \, \theta \int_{4\Delta^2_\theta}^{\Lambda_\theta^2} {dt}\frac{  1  }{4 \pi^2 v^3_F }\nonumber\\&&\,\frac{t - 4 \Delta_\theta^2 + 4 v_F^2 k_F^2}{ \sqrt{t-4 \Delta_\theta^2}\sqrt{t}}  \label{gapeq}
\end{eqnarray}
Recall that $\Delta_\theta = \Delta {\rm cos} \theta$ while the energy cutoff $\Lambda_\theta$ and the momentum cutoff ${\cal K}$ are related by expression
$\Lambda^2_\theta /4 = v_F^2 {\cal K}^2 + \Delta_\theta^2$ (integration is over momenta with $|k-k_F|< {\cal K}$). Actually, Eq. (\ref{gapeq}) is equivalent to the "gap" equation that relates the value of $\Delta$ with the coupling constants $g$, $g_m$ and the momentum cutoff $\cal K$.
 In the similar way
\begin{eqnarray}
&& 1/g - 1/g_m =  \int_{-1}^1 {\rm cos}^2 \, \theta d \, {\rm cos} \, \theta \int_{4\Delta^2_\theta}^{\Lambda_\theta^2}  {dt}\nonumber\\&&
 \frac{  1  }{4 \pi^2 v^3_F }\,\frac{t - 4 \Delta_\theta^2 + 4 v_F^2 k_F^2}{ \sqrt{t-4 \Delta_\theta^2}\sqrt{t}} \,\frac{ t-4 \Delta_\theta^2 }{t-E^2_{u,L=0,S=0}}\label{eq1}
\end{eqnarray}
Let us subtract Eq. (\ref{gapeq}) from Eq. (\ref{eq1}).
Assuming that $v_F k_F \gg v_F {\cal K} \gg \Delta$ we have:
\begin{eqnarray}
0 & = & \frac{  2k^2_F  }{ \pi^2 v_F } \int_{-1}^1 {\rm cos}^2 \, \theta d \, {\rm cos} \, \theta \int_{1}^{\infty}  {dz}\nonumber\\&&
\frac{1}{ \sqrt{z^2-1}}\,\frac{ E^2_{u,L=0,S=0}/(4 \Delta_\theta^2)-1}{z^2-E^2_{u,L=0,S=0}/(4\Delta^2_\theta)}\label{eqL0}
\end{eqnarray}
The integrals in this equation may be taken and the result is expressed through the hypergeometric functions:
\begin{eqnarray}
0 & = & \frac{ 4 k^2_F  }{ \pi^2 v_F } \Big[\frac{1}{4}w^4\sqrt{\pi} \Big(\frac{3}{8w}\pi^{3/2}-\frac{32}{15\sqrt{\pi}} F^{1/2, 1, 3}_{3/2, 7/2} (-w^2)\Big)\nonumber\\&&-\frac{1}{4} w^4 \sqrt{\pi} \Big(\frac{1}{2w} \pi^{3/2}-\frac{8}{3\sqrt{\pi}} F^{1/2, 1, 2}_{3/2, 5/2} ( -w^2)\Big)\nonumber\\&&+\frac{1}{3}w^2+\frac{1}{3} \Big]\label{rhs}
\end{eqnarray}
 where
 $$
 w = \frac{-i E_{u,L=0,S=0}}{2\Delta}
 $$
Technically we calculate the value of the integral in Eq. (\ref{eqL0}) at real values of $w$. Next, the obtained result is to be continued analytically to the whole complex plane. It is done in the way utilised inside the MAPLE package.

\begin{figure}[!thb]
\begin{center}
\includegraphics[scale=0.57,clip=true]{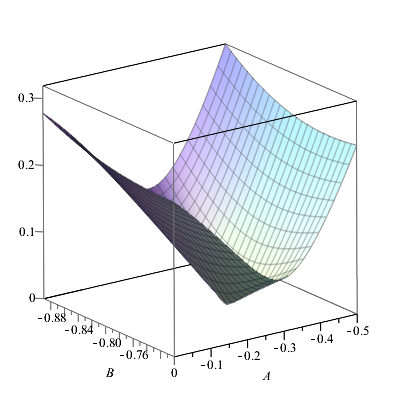}
\end{center}
\caption{\label{fig.3}  The absolute value of the right hand side of Eq.(\ref{rhs}) in the units of $\frac{ 4 k^2_F  }{ \pi^2 v_F }$ is represented as a function of $w = A+i B$.}
\label{fig:edge}
\end{figure}

Numerical solution of this equation gives
\begin{equation}
E_{u,S=0,L=0} = \sqrt{12/5} \, \Big( 1.007853779-0.3828669418\,i\Big) \, \Delta \end{equation}
This solution is illustrated by Fig. \ref{fig.3}, where the absolute value of the right hand side of Eq.(\ref{rhs}) in the units of $\frac{ 4 k^2_F  }{ \pi^2 v_F }$ is represented as a function of $w = A+i B$. One can see, that there is the solution in the physical part of the complex plane (at ${\rm Re}\,\omega < 0$, ${\rm Im}\,\omega < 0$). It corresponds to the energy gap of the given collective  mode.

{$L=0,S=1$.  We take components with $\alpha = 1,3$, $i=3$.}

In the $u$ - channel at $L=0, S=1$  the energy gap is equal to zero that leads to  the condition, which coincides with Eq. (\ref{gapeq}).
 In the similar way equation for the $v$ channel gives
\begin{equation}
E_{u,S=1,L=0} = 0, \, E_{v,S=1,L=0}  = E_{u,S=0,L=0}
   \end{equation}

{$L=1,S=0$.  We take components with $\alpha = 2$, $i=1,2$.}
In the $u$ channel
\begin{eqnarray}
\frac{1}{g} &=&  \int_{-1}^1 \frac{1-{\rm cos}^2 \, \theta}{2} d \, {\rm cos} \, \theta \int_{4\Delta^2_\theta}^{\Lambda_\theta^2}  {dt}
\frac{  1 }{4 \pi^2 v^3_F }\nonumber\\&&\frac{t - 4 \Delta_\theta^2 + 4 v_F^2 k_F^2}{ \sqrt{t-4 \Delta_\theta^2}\sqrt{t}} \,\frac{t-4 \Delta_\theta^2}{t-E^2_{u,L=1,S=0}}
\end{eqnarray}
At $E_{u,L=1,S=0}=0$ we may rewrite this equation in the form with the integration over $k$ instead of integration over $t$:
\begin{equation}
\frac{1}{g} = 8\pi \int \frac{d^3k}{(2\pi)^4}\frac{ {\rm sin}^2\,\theta {\cal M}^2(k)}{2(\Delta^2 {\rm cos}^2 \theta + {\cal M}^2(k))^{3/2}}\label{gapeq010}
\end{equation}
One can check that after the integration over $\theta$ the right hand sides of the two expressions Eq. (\ref{gapeq00}) and Eq. (\ref{gapeq010}) coincide.
Therefore, in the absence of the extra interaction that stabilizes direction of $\hat m$ in this channel the Goldstone boson appears as it should.

In the presence of this extra interaction we have the following equation for the determination of $E_{u,L=1,S=0}$:
\begin{eqnarray}
\frac{1}{g_m} &=&  \int_{-1}^1 \frac{1-{\rm cos}^2 \, \theta}{2} d \, {\rm cos} \, \theta \int_{4\Delta^2_\theta}^{\Lambda_\theta^2}  {dt}
\frac{  1 }{4 \pi^2 v^3_F }\nonumber\\&&\frac{t - 4 \Delta_\theta^2 + 4 v_F^2 k_F^2}{ \sqrt{t-4 \Delta_\theta^2}\sqrt{t}} \,\frac{\Big( t-4 \Delta_\theta^2 \Big)E_{u,L=1,S=0}^2}{t(t-E^2_{u,L=1,S=0})}
\end{eqnarray}
In the $v$ channel we have
\begin{eqnarray}
\frac{1}{g} &=&  \int_{-1}^1 \frac{1-{\rm cos}^2 \, \theta}{2} d \, {\rm cos} \, \theta \int_{4\Delta^2_\theta}^{\Lambda_\theta^2}  dt
\frac{  1 }{4 \pi^2 v^3_F }\nonumber\\&&\frac{t - 4 \Delta_\theta^2 + 4 v_F^2 k_F^2}{ \sqrt{t-4 \Delta_\theta^2}\sqrt{t}} \,\frac{t}{t-E^2_{v,L=1,S=0}}
\end{eqnarray}
Subtracting the gap equation we may represent this expression as follows \begin{eqnarray}
&& \frac{1}{g_m} =  \int_{-1}^1 \frac{1-{\rm cos}^2 \, \theta}{2} d \, {\rm cos} \, \theta \int_{4\Delta^2_\theta}^{\Lambda_\theta^2}  {dt}
\frac{  1  }{4 \pi^2 v^3_F }\nonumber\\&&\frac{t - 4 \Delta_\theta^2 + 4 v_F^2 k_F^2}{ \sqrt{t-4 \Delta_\theta^2}\sqrt{t}} \,\frac{\Big( E^2_{v,L=1,S=0}(t-4 \Delta_\theta^2) + 4 \Delta_\theta^2 t \Big)}{t (t-E^2_{v,L=1,S=0})}
\end{eqnarray}
The value of $1/g_m$ should be sufficiently large in order to make vacuum stable. The critical value  $g^{(c)}_m$ is determined by equation:
\begin{eqnarray}
&&\frac{1}{g^{(c)}_m} =  \int_{-1}^1 \frac{1-{\rm cos}^2 \, \theta}{2} d \, {\rm cos} \, \theta \int_{4\Delta^2_\theta}^{\Lambda_\theta^2}  {dt}
\frac{  1  }{4 \pi^2 v^3_F }\nonumber\\&&\frac{t - 4 \Delta_\theta^2 + 4 v_F^2 k_F^2}{ \sqrt{t-4 \Delta_\theta^2}\sqrt{t}} \,\frac{ 4 \Delta_\theta^2 }{t }=\frac{ 2 k^2_F  }{3\pi^2 v_F }
\end{eqnarray}
At this critical value of $g_m$ the energy gap $E_{v,L=1,S=0}$ is close to zero.
We get
\begin{eqnarray}
&&-1/g_m + 1/g_m^{(c)}= \frac{ 2 k^2_F  }{\pi^2 v_F } \int_{-1}^1 \frac{1-x^2}{2} dx \nonumber\\&&\int_{1}^{\infty}  {dz}
\,\frac{1}{ \sqrt{z^2-1}} \,\frac{ w^2+x^2 }{x^2 z^2 + w^2 }, \quad w = - \frac{i E_{u,L=1,S=0}}{2\Delta}
\end{eqnarray}
and
\begin{eqnarray}
&& -1/g_m + 1/g_m^{(c)}= \frac{ 2 k^2_F  }{\pi^2 v_F } \int_{-1}^1 \frac{1-x^2}{2} dx \nonumber\\&&\int_{1}^{\infty}  {dz}
\,\frac{1}{ \sqrt{z^2-1}} \,\frac{ w^2 }{x^2 z^2 + w^2 },\quad w = - \frac{i E_{v,L=1,S=0}}{2\Delta}\label{Ev}
\end{eqnarray}
The integration gives correspondingly
\begin{eqnarray}
 0 &= & 1/g_m - 1/g_m^{(c)} \nonumber\\&&+ \frac{ 4 k^2_F  }{ \pi^2 v_F } \Big[\frac{1}{16}w^4\sqrt{\pi} \Big(\frac{1}{4w}\pi^{3/2}-\frac{16}{15\sqrt{\pi}} F^{1/2, 1, 2}_{3/2, 7/2} (-w^2)\Big)\nonumber\\&&+\frac{1}{16} w^4 \sqrt{\pi} \Big(\frac{1}{w} \pi^{3/2}-\frac{8}{3\sqrt{\pi}} F^{1/2, 1, 1}_{3/2, 5/2} ( -w^2)\Big)\nonumber\\&&-\frac{1}{6}w^2+\frac{1}{3} \Big]
\end{eqnarray}
for the $u$ - mode
and
\begin{eqnarray}
 0 &=& 1/g_m - 1/g_m^{(c)}\nonumber\\&& + \frac{ 4 k^2_F  }{ \pi^2 v_F } \Big[\frac{1}{8}w^4\sqrt{\pi} \Big(\frac{1}{2w}\pi^{3/2}-\frac{8}{3\sqrt{\pi}} F^{1/2, 1, 2}_{3/2, 5/2} (-w^2)\Big)\nonumber\\&&+\frac{1}{8} w^4 \sqrt{\pi} \Big(\frac{1}{w} \pi^{3/2}-\frac{4}{\sqrt{\pi}} F^{1/2, 1, 1}_{3/2, 3/2} ( -w^2)\Big)\nonumber\\&&-\frac{1}{2}w^2 \Big]\label{rhs2}
\end{eqnarray}
for the $v$ - mode.

It appears, that for $1/g_m^{(c)} > 0 > 1/g_m $ the first equation has the solution  for real value of $w$ and imaginary value of $E_{u,L=1,S=0}$. For $0 = 1/g_m$ the solution with $E_{u,L=1,S=0}=0$ appears, while for  $0 < 1/g_m$ there are no solutions of this equation in the physical region of $\omega$.  (For ${\rm Im}\,\omega = 0$ the physical region is ${\rm Re}\,\omega \ge 0$.)

The second equation for $1/g_m^{} < 1/g_m^{(c)}$ has the solution with real $w$ and pure imaginary $E_{v,L=1,S=0}$, as it was pointed out above. For $1/g_m^{(c)} = 1/g_m$ the solution with $E_{v,L=1,S=0}=0$ appears, while for  $1/g_m^{(c)} < 1/g_m$ there is the solution with real negative $w$. It does not represent any solution of the original equation given by the integral and therefore belongs to the unphysical region of $w$.

\begin{figure}[!thb]
\begin{center}
\includegraphics[scale=0.57,clip=true]{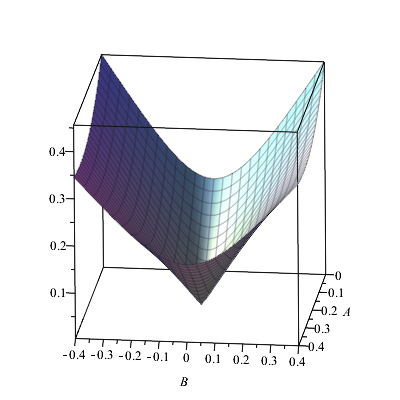}
\end{center}
\caption{\label{fig.4}  The absolute value of the right hand side of Eq.(\ref{rhs2}) in the units of $\frac{ 4 k^2_F  }{ \pi^2 v_F }$ is represented as a function of $w = A+i B$ for $1/g-1/g_m = -0.2 \frac{ 4 k^2_F  }{ \pi^2 v_F }$. }
\label{fig:edge}
\end{figure}

This situation is illustrated by Fig. \ref{fig.4}, where the absolute value of the right hand side of Eq.(\ref{rhs2}) in the units of $\frac{ 4 k^2_F  }{ \pi^2 v_F }$ is represented as a function of $w = A+i B$ for $1/g_m-1/g^{(c)}_m = -0.2 \frac{ 4 k^2_F  }{ \pi^2 v_F }$. One can see, that there is the solution at ${\rm Im}\,\omega = 0$, ${\rm Re}\,\omega > 0$). It corresponds to the pure imaginary energy gap, and indicates the instability of vacuum. When the value of $1/g_m-1/g^{(c)}_m$ is increased, the solution approaches zero. At $1/g_m-1/g^{(c)}_m > 0$ the solution of Eq. (\ref{rhs2}) exists at real negative values of $\omega$ that are not physical because they do not correspond to any solutions of Eq. (\ref{Ev}).

{$L=1,S=1$. We take components with $\alpha = 1,3$; $i=1,2$.}
It appears, that here the equations for the determinations of the gaps are the same as for $L=1,S-0$ with the modes $u$ and $v$ exchanged.
We come to
\begin{equation}
E_{u,S=1,L=1} = E_{v,S=0,L=1}, \quad E_{v,S=1,L=1} = E_{u,S=0,L=1}
\end{equation}


One can see, that in the channels with $L=0$, where the gaps of the order of $\Delta$ appear, these gaps satisfy the Nambu sum rule
$$
E^2_u+ E_v^2 = 4 \langle \Delta_\theta^2\rangle = \frac{12}{5} \Delta^2
$$

We come to the conclusion, that vacuum becomes stable for $g_m<g_m^{(c)}$, but the Higgs modes in the channels with $L=1$ do not exist.


\begin{thebibliography}{99}

\bibitem{Dmitriev2012}
R. Sh. Askhadullin, V. V. Dmitriev, D. A. Krasnikhin, P. N. Martynov, A. A. Osipov, A. A. Senin, A. N. Yudin,
Phase diagram of superfluid $^3$He in "nematically ordered" aerogel,
JETP Lett. {\bf 95}, 326 (2012).

\bibitem{Dmitriev2014}
R.Sh. Askhadullin, V.V.Dmitriev, P.N. Martynov, A.A. Osipov, A.A. Senin,  A.N. Yudin,
Anisotropic 2D Larkin  Imry  Ma state in polar distorted ABM phase of $^3$He in "nematically ordered" aerogel,
Pis'ma ZhETF, {\bf 100},  747--753 (2014);
arXiv:1410.5194.

\bibitem{Dmitriev2}
V. V. Dmitriev, A. A. Senin, A. A. Soldatov, and A. N. Yudin, Phys. Rev .Lett. 115, 165304 (2015).

\bibitem{Mineev2014}
V.P. Mineev,
Half-quantum vortices in polar phase of superfluid $^3$He,
J. Low Temp. Phys. {\bf 177},   48--58 (2014);
arXiv:1402.2111.

\bibitem{aerogel0}
K. Aoyama and R. Ikeda, Phys. Rev. B 73, 060504 (2006).

\bibitem{Volovik2003}
G.E. Volovik,
{\it The Universe in a Helium Droplet},
Clarendon Press,  Oxford (2003).

\bibitem{VollhardtWolfle1990}
D. Vollhardt  and P.  W\"olfle,
{\it The superfluid phases of helium 3},  Taylor and Francis, London
(1990).



\bibitem{He3}
V. Alonso and V. N. Popov,
Functional for the hydrodynamic action and the Bose
spectrum of superfluid Fermi systems of the He3 type,
Zh. Eksp. Teor. Fiz. {\bf 73}, 1445--1459 (1977).


\bibitem{He3gauss}
P. N. Brusov and V. N. Popov,
Stability of the Bose spectrum of superfluid systems of the He3 type,
Zh. Eksp. Teor. Fiz. {\bf 78}, 234--245 (1980).

\bibitem{He3B}
P.N. Brusov and V.N. Popov,
Nonphonon branches of the Bose spectrum in the B phase of systems of the He3 type,
JETP  {\bf 51}, 1217--1222 (1980).

\bibitem{BrusovPopov1980}
P.N. Brusov and V.N. Popov,
"Zero-phonon branches of the Bose spectrum in the A phase of a system of the He3 type",
JETP {\bf 52}, 945--949 (1980).

\bibitem{Brusovs}
Peter Brusov, Pavel Brusov, "Collective Excitations in
Unconventional
Superconductors
and Superfluids", World Scientific Publishing Co. Pte. Ltd. (2010)


\bibitem{VolovikVachaspati1996}
G.E. Volovik and T. Vachaspati,
"Aspects of $^3$He and the  standard electroweak model,"
Int. J. Mod. Phys. {\bf B~10}, 471--521 (1996);
cond-mat/9510065.


\bibitem{VolovikKhazan1982}
G.E. Volovik, M.V. Khazan,
 "Dynamics of the A-phase of  $^3$He at low pressure,"
JETP {\bf 55},  867--871 (1982);
"Classification of the collective modes of the order parameter in superfluid $^3$He," JETP {\bf 58}, 551--555  (1983).

\bibitem{Volovik1990}
G.E. Volovik,
"Symmetry in superfluid $^3$He",
  in: {\bf Helium Three}, eds. W.P.Halperin, L.P.Pitaevskii, Elsevier Science
   Publishers B.V., pp. 27--134 (1990).

\bibitem{Nambu1985}
Yoichiro Nambu,
Fermion - boson relations in BCS type theories,
Physica D {\bf 15},  147--151 (1985);
Energy gap, mass gap, and spontaneous symmetry breaking,
in: {\it BCS: 50 Years}, eds. L.N. Cooper and D. Feldman, World Scientific
(2010).

\bibitem{VZ2015}
  G.~E.~Volovik and M.~A.~Zubkov,
  ``Scalar excitation with Leggett frequency in $^3$He-B and the 125 GeV Higgs particle in top quark condensation models as pseudo-Goldstone bosons,''
  Phys.\ Rev.\ D {\bf 92}, no. 5, 055004 (2015)
  doi:10.1103/PhysRevD.92.055004
  [arXiv:1410.7097 [hep-ph]].

\bibitem{VolovikZubkov2014}
G.E. Volovik and M.A. Zubkov,
Higgs bosons in particle physics and in condensed matter,
J. Low Temp. Phys. {\bf 175}, 486--497 (2014).




\bibitem{VolovikZubkovHiggs}
G.E. Volovik and M.A. Zubkov,
The Nambu sum rule and the relation between the masses of composite Higgs
bosons,
Phys. Rev. D  {\bf 87}, 075016 (2013);
Nambu sum rule in the NJL models: from superfluidity  to the models
of top quark condensation,
Pis'ma ZhETF {\bf 97},  344--349 (2013); JETP Lett. {\bf 97},  301--306
(2013).





\end{thebibliography}
\end{document}